# Wave propagation in one-dimensional nonlinear acoustic metamaterials

Xin Fang[1], Jihong Wen[1], Bernard Bonello[2], Jianfei Yin[1] and Dianlong Yu[1]

[1]*Laboratory of Science and Technology on Integrated Logistics Support, National University of Defense Technology, Changsha 410073, Hunan, China*

[2]*Institut des NanoSciences de Paris (INSP-UMR CNRS 7588), Université Pierre et Marie Curie (box 840) 4, place Jussieu, 75252 Paris Cedex 05, France*

**E-mail:** wenjihong@vip.sina.com; nmhsyjf@hotmail.com

**Abstract**

The propagation of waves in the nonlinear acoustic metamaterials (NAMs) is fundamentally different from that in the conventional linear ones. In this article we consider two one-dimensional NAM systems featuring respectively a diatomic and a tetratomic meta unit-cell. We investigate the attenuation of the wave, the band structure and the bifurcations to demonstrate novel nonlinear effects, which can significantly expand the bandwidth for elastic wave suppression and cause nonlinear wave phenomena. Harmonic averaging approach, continuation algorithm, Lyapunov exponents are combined to study the frequency responses, the nonlinear modes, bifurcations of periodic solutions and chaos. The nonlinear resonances are studied and the influence of damping on hyper-chaotic attractors is evaluated. Moreover, a "quantum" behavior is found between the low-energy and high-energy orbits. This work provides an important theoretical base for the further understandings and applications of NAMs.

## 1. Introduction

Like the phononic crystals [1,2], acoustic metamaterials [3-6] (AMs) are artificial medias that gain their properties from structure rather than composition. They generally feature built in resonators with subwavelength dimensions, rendering the concepts of effective mass density and effective elastic constants relevant for characterizing them. Because of their ability to manipulate the propagation of sound, they have become a hot topic during the last decade [7]. By now, most studies [8-12] in this field are dealing with linear AMs (LAMs) which are based on a locally resonant (LR) mechanism [1] to yield negative indexes, super-lensing effects [13], or wave guiding. Different from the Bragg mechanism, LR mechanism paves the way to manipulate low-frequency waves with small "meta-atoms" [13]. However, the low-frequency LR bandgap is generally narrow [7] and heavy "atoms" are necessary to enlarge this bandgap.

The concept of nonlinear metamaterial was first introduced in electromagnetism to investigate the photonic metamaterials in 2003 [14]. According to particular needs, a nonlinear response can be constructed on purpose and even enhanced as compared to conventional nonlinearities found in natural materials, such as tenability, electromagnetically induced transparency, plasmonics, active media, etc [15,16]. In contrast, it is only recently that nonlinear acoustic metamaterials (NAMs) have emerged.

Compared with NAMs, studies on nonlinear periodic structures [17] (NPSs) have a longer history that dates back to 1983 when Nesterenko highlighted the soliton in a granular crystal [18]. Since then, most of the investigations in the field are dealing with granular crystals interacting in a nonlinear way through Hertzian contact [19,20]. Bistable lattices (another NPS) are studied recently [21]. Actually, NPSs still attract much attention and even they got an intense development since 2006 [19, 21-23]. The presence of nonlinear media in periodic structure can be utilized to realize interesting devices, such as acoustic diodes [24-26] and acoustic lenses [27]. New physical properties that are different from the linear phononic crystals are found, such as unidirectional transition [21], waves coupling [33], subharmonic frequencies [34,35] , discrete breathers [23], soliton waves [36,37] and surface waves [38]. Both simulations and experiments demonstrate that bifurcations may be observed in granular crystals [22,26,39]. Using a perturbation approach and the harmonic balance method, Narisetti *et al.* [40,41] recently studied the amplitude-dependent dispersions, stop band properties, and wave beaming in nonlinear periodic granular media. On their side, Manktelow *et al.*[42] investigated the wave propagation in layered NPS based on a perturbation approach. Furthermore, experimental works highlighted the role played by the critical amplitude in energy transmission [43] and bifurcation-induced bandgap reconfiguration [44] in one-dimensional (1D) NPSs. Recently, a 1D strong NPS has been proved to increase the velocity of sound and therefore the acoustic impedance [45]. However, while the granular crystals are suitable for ultrasonic applications, it is hard to consider them at low frequency regime because of the high contact stiffness they inherently feature.

Because of their promising applications, AMs with both low-frequency and broadband properties, attract currently much attention. However, the mechanisms for both properties simultaneously happen are difficult to realize. LAMs consist of linear "meta-atom", but when the linear "meta-atom" gets nonlinear in NAM, properties of the wave propagation get different patterns. In the recent works [46][47], the wave propagation in diatomic and tetratomic NAMs are analyzed using the homotopy analysis method, and we found that the chaotic bands resulting from bifurcations can significantly enlarge the width of the forbidden bands. This finding demonstrates that chaos is a novel and promising mechanism to achieve simultaneously low-frequency and broad band in both mono-bandgap NAMs and multi-bandgap NAMs and still that a strong nonlinearity is beneficial to expand the bandwidth by several times.

However, there are lots of phenomena arising in NAMs that have not yet been fully explained nor demonstrated. For example, why are the responses in the first pass band similar to ones observed with LAMs? Under which conditions can the elastic energy propagate in the bandgap? When will the wave be amplified by chaos? In the tetratomic system, why the nonlinearity has larger influence on the nonlinear LR bandgaps than it has for the Bragg gaps? Furthermore, some problems remain still unexplored, for example, the influence of the nonlinearity on the structure of the bands, the features of chaos and their differences, the jumps, the nonlinear resonances and their stabilities and the influence of the damping on chaos.

In this paper we attempt to answer these questions with the help of the frequency response analysis, the bifurcation theories, the Lyapunov exponents (LEs) and the fractal dimensions.

## 2. Models

### 2.1. Diatomic model

The 1D damped diatomic nonlinear metamaterial model is illustrated in figure 1. The nonlinear "meta-atom" of this periodic model consists of a linear oscillator $m$ and a damped nonlinear Duffing oscillator $m_r$ with cubic nonlinear stiffness $k_1\Delta+k_2\Delta^3$, where $\Delta$ denotes the relative displacement and $k_1$ ($k_2$) symbolizes the linear (nonlinear) stiffness. In addition, a linear viscous damping term $c \cdot \dot{\Delta}$ is taken into consideration.

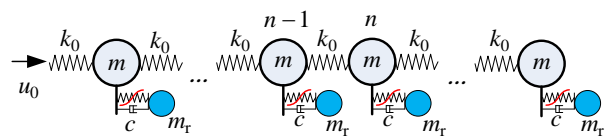

**Figure 1.** Model of the diatomic NAM

Defining $x_n$ and $y_n$ as displacements of the linear and nonlinear oscillators in the $n$th cell, respectively; the differential equation for the $n$th cell reads

$$\begin{cases} \ddot{x}_n = \beta_0(x_{n+1} + x_{n-1} - 2x_n) + \eta\left[\zeta(\dot{y}_n - \dot{x}_n) + \beta_1(y_n - x_n) + \beta_2(y_n - x_n)^3\right] \\ \ddot{y}_n = -\zeta(\dot{y}_n - \dot{x}_n) - \beta_1(y_n - x_n) - \beta_2(y_n - x_n)^3 \end{cases} \quad (1)$$

where the dot denotes the derivation with respect to time $t$. The definitions of the generalized parameters are: mass ratio $\eta = m_r/m$, stiffness ratio $\beta_0 = k_0/m = \omega_s^2$, $\beta_1 = k_1/m_r = \omega_r^2$, $\beta_2 = k_2/m_r$. The generalized frequency is $\Omega = \omega/\omega_s$ and the damping coefficient is $\zeta = c/m_r$. The strength factor of nonlinearity is defined as $\sigma = 3\beta_2 A_0^2/\beta_1$, where $A_0$ stands for the amplitude of the incident wave $u_0 = A_0 \sin(\omega t)$ at the left end. It is a weak nonlinearity if $\sigma \ll 1$. The values of the parameters used in this model are: $m=1$, $m_r=0.5$, $\beta_0=10\pi$, $\beta_1=15\pi$, $\beta_2=10^5$. If $A_0=0.005$, $\sigma=0.16$, it is moderate nonlinearity.

Applying the Bloch theorem for periodic structure, the displacement reads $\mathbf{u}(r+a) = \mathbf{u}(r)\exp(i\kappa a)$, where $\kappa$ is the wave vector and $a$ is the lattice constant. We can solve the dispersion relations with perturbation approach, harmonic balance method or homotopy analysis method. The algorithms for these approaches are described in [47]. The dispersion solutions are eigen waves in the infinite metamaterial without external input.

We have considered a diatomic NAM with a finite length: 8 unit cells are involved in the model. Actually, further increasing the number of cells hardly changes the intrinsic characteristics of the NAM but would decrease the accuracy of the bifurcation diagram. Therefore, there are 16 degrees-of-freedom (DoF) and the state space has 32 dimensions (32D, $n$D symbolizes $n$-dimensional). In practice, applications would imply the dynamic of waves in a finite system with external excitations. The monochromatic sinusoidal wave $u_0(t)$ drives the system from the left end. The right end is free and the terminal linear oscillator is analyzed.

Beside the diatomic model, tetratomic model is also discussed to figure out the influences on multiple bandgaps.

**2.1. Tetratomic model**

The model of the tetratomic NAM is illustrated in figure 2. A "molecule" of this model is composed of three identical linear "atoms" and one nonlinear "atom". Their parameters are labeled in the figure, where $x_n$, $y_n$, $z_n$, $r_n$ denote the displacements of corresponding oscillators in the $n$th cell, respectively. The tetratomic NAM has more complicated dynamics.

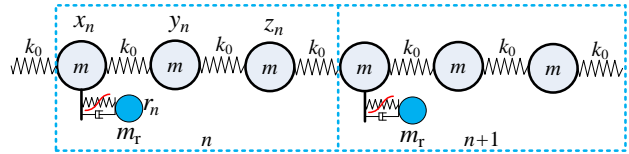

**Figure 2.** Model of the tetratomic NAM

The generalized motion equation of the $n$th cell can be simplified as

$$\begin{cases} \ddot{x}_n = \beta_0(y_n + z_{n-1} - 2x_n) + \eta\left[\zeta(\dot{r}_n - \dot{x}_n) + \beta_1(r_n - x_n) + \beta_2(r_n - x_n)^3\right] \\ \ddot{y}_n = \beta_0(x_n + z_n - 2y_n) \\ \ddot{z}_n = \beta_0(x_{n+1} + y_n - 2z_n) \\ \ddot{r}_n = -\zeta(\dot{r}_n - \dot{x}_n) - \beta_1(r_n - x_n) - \beta_2(r_n - x_n)^3 \end{cases} \quad (2)$$

The definitions of other parameters in Eq. (2) are identical with those in Eq. (1)

The algorithms to analyze the bifurcation and characteristics of chaos in the tetratomic model are also identical with the diatomic model. The difference is the chain we considered here has only four cells, so the state space is also 32D. The parameters of the model are: $m=1$, $m_r=2$, $\eta=2$, $\beta_0=225=\omega_s^2$, $\beta_1=25=\omega_{sr}^2$, $\beta_2=5\times10^4$, and $\Omega=\omega/\omega_s$. $\omega_s$ is 3 times of $\omega_{rs}$, which is similar with the continuous metamaterials which requires soft resonators. A larger but still weak damping $\zeta=0.02$ is adopted. Assuming that the driving amplitude is $A_0=0.01$ and the nonlinear strength is $\sigma=0.6$, so it is strongly nonlinear system in this case.

## 3. Theories

### 3.1. Method in frequency domain

(1) Some Algebraic laws

The analysis of the nonlinear dynamics of high dimensional systems generally involves high order operations on matrices and vectors. For convenience, let us define at first some algebraic operations.

We define the "element product" of two $n$D columns vectors $\mathbf{x}=[x_1\ x_2\cdots x_n]^T$ and $\mathbf{y}=[y_1\ y_2\cdots y_n]^T$ to be a $n$D vector $\mathbf{z}$, components of which are:

$$\mathbf{z} = \mathbf{x}\circ\mathbf{y} = [\cdots x_i y_i \cdots]^T, \text{i.e., } z_i = x_i y_i \quad (3)$$

The element product of two vectors can be abbreviated as $\mathbf{xy}$. Therefore, the notation for the $m$th power of $\mathbf{x}$ is $\mathbf{x}^m = [\cdots x_i^m \cdots]^T$. It can be easily established that the element product is commutative $\mathbf{xy}=\mathbf{yx}$, distributive $(\mathbf{x}+\mathbf{y})\mathbf{z}=\mathbf{xz}+\mathbf{yz}$, and associative $\mathbf{xyz}=\mathbf{zyx}$.

Similarly, we define the element product of two matrices as being $\mathbf{A}\circ\mathbf{B}=[a_{ij}b_{ij}]$, where the dot '∘' cannot be omitted. It is straightforward to show the element product of two matrices obeys the commutative, distributive and associative laws as well. However, the product between a matrix and an element product of two vectors does generally not observe the associative law, *i.e.* $\mathbf{A}(\mathbf{x}\circ\mathbf{y})\ne(\mathbf{Ax})\circ\mathbf{y}$, which could become an issue in practice. To overcome this obstacle and to ensure that this operation gets associative, we set the following transformation

$$\mathbf{A}(\mathbf{x}\circ\mathbf{y}) = (\mathbf{A}\circ[[\mathbf{x}]])\mathbf{y} \quad (4)$$

where the notation $[[\mathbf{x}]]$ stands for the square matrix built upon the vector $\mathbf{x}$, namely

$$[[\mathbf{x}]] = [\underbrace{\mathbf{x}\ \mathbf{x}\ \cdots\ \mathbf{x}}_{n}]^T$$

With this definition, together with the commutativity, it comes $(\mathbf{A}\circ[[\mathbf{x}]])\mathbf{y}=(\mathbf{A}\circ[[\mathbf{y}]])\mathbf{x}$. Moreover, it can be easily shown that the partial derivative of the element product is:

$$\frac{\partial\mathbf{A}(\mathbf{x}\circ\mathbf{y})}{\partial\mathbf{y}} = \mathbf{A}\circ[[\mathbf{x}]], \quad \frac{\partial\mathbf{A}(\mathbf{x}^m)}{\partial\mathbf{x}} = m\mathbf{A}\circ[[\mathbf{x}^{m-1}]] \quad (5)$$

These algebraic rules allow for a convenient analysis of the high-dimensional nonlinear equations and lead to compact formulas.

(2) Frequency responses

Achieving the analytical solutions of Eqs. (1) and (2) for a high-dimensional system is not an easy task. However, the frequency responses can be approximated with the help of the numerical integral method, the averaging method, or the harmonic balance method (HBM).

The latter can be implemented to find the approximate steady frequency responses. To this end, we define the coordinate transformation $\hat{y}_n = y_n - x_n$ in Eq.(1) or $\hat{y}_n = r_n - x_n$ in Eq. (2). With this transformation, the equation of motion for the finite nonlinear metamaterial model reads

$$\mathbf{M}\ddot{\mathbf{y}} + \mathbf{C}\dot{\mathbf{y}} + \mathbf{K}\mathbf{y} + \mathbf{N}\mathbf{y}^3 = \mathbf{f}\cos\omega t \tag{6}$$

where $\mathbf{M}$, $\mathbf{C}$, $\mathbf{K}$ and $\mathbf{N}$ are mass, damping, stiffness and nonlinear coefficient matrices of the whole transformed system; $\mathbf{f}$ stands for the node force vector applied on every masses.

Let us assume that the steady response has the form $\mathbf{y} = \mathbf{a}\cos\omega t + \mathbf{b}\sin\omega t$, where $\mathbf{a}$ and $\mathbf{b}$ are constant vectors. The first-order HBM procedure leads to a system of algebraic equations:

$$\begin{cases} [\mathbf{K} - \omega^2\mathbf{M}]\mathbf{a} + \omega\mathbf{C}\mathbf{b} + 3\mathbf{N}\left((\mathbf{a}^2 + \mathbf{b}^2)\mathbf{a}\right)/4 = \mathbf{f} \\ [\mathbf{K} - \omega^2\mathbf{M}]\mathbf{b} - \omega\mathbf{C}\mathbf{a} + 3\mathbf{N}\left((\mathbf{a}^2 + \mathbf{b}^2)\mathbf{b}\right)/4 = \mathbf{0} \end{cases} \tag{7}$$

The solutions of Eq. (7) would accurately describe the responses $\mathbf{Y} = \sqrt{\mathbf{a}^2 + \mathbf{b}^2}$ to the driving force, of the metamaterial. However, HBM method does not directly allow investigating the stability of the solutions. This is the reason why we adopted in this work the harmonic average approach (HAA) [48][49] instead, to study the frequency responses. Within this approach, the solution is assumed to have the form:

$$\begin{cases} \mathbf{y} = \mathbf{u}(t)\cos\theta + \mathbf{v}(t)\sin\theta \\ \dot{\mathbf{y}} = -\omega\mathbf{u}(t)\sin\theta + \omega\mathbf{v}(t)\cos\theta \end{cases} \tag{8}$$

where $\theta = \omega t$. The derivatives with respect to time $t$ of the formula in (8) are

$$\begin{cases} \dot{\mathbf{y}} = (\dot{\mathbf{u}} + \omega\mathbf{v})\cos\theta + (\dot{\mathbf{v}} - \omega\mathbf{u})\sin\theta \\ \ddot{\mathbf{y}} = (\omega\dot{\mathbf{v}} - \omega^2\mathbf{u})\cos\theta - (\omega\dot{\mathbf{u}} + \omega^2\mathbf{v})\sin\theta \end{cases} \tag{9}$$

Comparing the expressions of $\dot{\mathbf{y}}$ in Eqs. (8) and (9), we obtain

$$\omega\mathbf{M}(\dot{\mathbf{u}}\cos\theta + \dot{\mathbf{v}}\sin\theta) = \mathbf{0} \tag{10}$$

The further substitution of (8) and (9) into (6) gives another form of the equation of motion:

$$\omega\mathbf{M}\left(\dot{\mathbf{v}}\cos\theta - \dot{\mathbf{u}}\sin\theta\right) + [\mathbf{K} - \omega^2\mathbf{M}](\mathbf{u}\cos\theta + \mathbf{v}\sin\theta) + \mathbf{C}\left((\dot{\mathbf{u}} + \omega\mathbf{v})\cos\theta + (\dot{\mathbf{v}} - \omega\mathbf{u})\sin\theta\right) \\ + \mathbf{N}\{(\mathbf{u}^3 - 3\mathbf{u}\mathbf{v}^2)\cos^3\theta + (\mathbf{v}^3 - 3\mathbf{u}^2\mathbf{v})\sin^3\theta + 3\mathbf{u}^2\mathbf{v}\sin\theta + 3\mathbf{u}\mathbf{v}^2\cos\theta\} = \mathbf{f}\cos\theta \tag{11}$$

Then, assuming that $\mathbf{u}$ and $\mathbf{v}$ are constant, by calculating $(11)\times\sin\theta - (10)\times\cos\theta$ and integrating the result from 0 to $2\pi$, we obtain

$$2\omega\dot{\mathbf{u}} = \mathbf{M}^{-1}\left[(\mathbf{K} - \omega^2\mathbf{M})\mathbf{v} - \omega\mathbf{C}\mathbf{u} + 3\mathbf{N}\left((\mathbf{v}^2 + \mathbf{u}^2)\mathbf{v}\right)/4\right] \tag{12}$$

Similarly, calculating $(11)\times\cos\theta + (10)\times\sin\theta$ and integrating over the interval $[0, 2\pi]$ gives:

$$2\omega\dot{\mathbf{v}} = -\mathbf{M}^{-1}\left[(\mathbf{K} - \omega^2\mathbf{M})\mathbf{u} + \omega\mathbf{C}\mathbf{v} + 3\mathbf{N}\left((\mathbf{v}^2 + \mathbf{u}^2)\mathbf{u}\right)/4 - \mathbf{f}\right] \tag{13}$$

The steady solutions correspond to the condition $\dot{\mathbf{u}} = \dot{\mathbf{v}} = \mathbf{0}$ and their expressions are therefore:

$$\begin{cases} (\mathbf{K} - \omega^2\mathbf{M})\mathbf{v} - \omega\mathbf{C}\mathbf{u} + 3\mathbf{N}\left((\mathbf{v}^2 + \mathbf{u}^2)\mathbf{v}\right)/4 = 0 \\ (\mathbf{K} - \omega^2\mathbf{M})\mathbf{u} + \omega\mathbf{C}\mathbf{v} + 3\mathbf{N}\left((\mathbf{v}^2 + \mathbf{u}^2)\mathbf{u}\right)/4 = \mathbf{f} \end{cases} \tag{14}$$

The amplitude of the response is $\mathbf{Y} = \sqrt{\mathbf{u}^2 + \mathbf{v}^2}$.

The derivation of these solutions using HAA turns the steady response problem into an equilibrium problem of differential equations. Actually, the solutions in Eqs. (14) and (7) are equal. However, HAA allows for analyzing the stability of the solutions of both the systems of differential equations (12) and (13) via the Jacobian matrix. Omitting the terms in $2\omega \neq 0$ in the left sides of (12) and (13) since they will not influence the properties of the solutions, the Jacobian matrix $\mathbf{J}$ is derived by performing the derivations of (12) and (13) with respect to the vectors $\mathbf{u}$ and $\mathbf{v}$, namely

$$\mathbf{J} = \begin{bmatrix} \mathbf{M}^{-1} & \mathbf{0} \\ \mathbf{0} & -\mathbf{M}^{-1} \end{bmatrix} \begin{bmatrix} -\omega\mathbf{C} + 3\mathbf{N} \circ [[\mathbf{uv}]]/2 & \mathbf{K} - \omega^2\mathbf{M} + 3\mathbf{N} \circ [[3\mathbf{v}^2 + \mathbf{u}^2]]/4 \\ \mathbf{K} - \omega^2\mathbf{M} + 3\mathbf{N} \circ [[3\mathbf{u}^2 + \mathbf{v}^2]]/4 & \omega\mathbf{C} + 3\mathbf{N} \circ [[\mathbf{uv}]]/2 \end{bmatrix} \quad (15)$$

Therefore, with the solutions coming from Eq. (15), their stability can be determined: if the real part of an eigenvalue of $\mathbf{J}$ is positive, the corresponding steady solution is unstable; if a real eigenvalue goes from negative to positive, saddle-node (SN) bifurcation occurs, and if a pair of conjugate complex eigenvalues crosses the imaginary axis, Hopf bifurcation appears [50]. The succinct expressions derived above are universal for the NAMs and differential equations with cubic nonlinearity.

The pseudo-arclength continuation algorithm is further adopted to solve Eq. (14). Within this frame, Newton method (MATLAB *fsolve*) is used to find the solutions. Let us notice that if $\mathbf{C}=\mathbf{0}$, this formula reveals that $\mathbf{v}$ must read $\mathbf{v}=\mathbf{0}$.

### 3.2. Bifurcation analysis method in time domain

From now on, the periodic solutions will be mostly derived using time domain methods, which are parts of software such as AUTO [51]. Within this kind of approach, the integrations of the equations of motion are made in the state space. Furthermore, the method allowing to derive the LEs [52][53] is also based on numerical integrations.

The incident elastic wave transforms the model in a non-autonomous system. However, when calculating the spectra of LEs and bifurcation diagrams, one must transform the non-autonomous system to an autonomous one. To calculate the former, the input wave is expressed as

$$u_0 = A_0 \sin\theta, \quad \dot{\theta} = \omega, \quad \theta(0) = 0 \quad (16)$$

*i.e.* it becomes a boundary value problem. Therefore, it becomes a 33D system in this situation.

When calculating the bifurcation diagrams, a differential system is added.

$$\begin{cases} \dot{\phi} = \phi + \omega\varphi - \phi(\phi^2 + \varphi^2) \\ \dot{\varphi} = -\omega\phi + \varphi - \varphi(\phi^2 + \varphi^2) \end{cases} \quad (17)$$

This system has a unique asymptotically stable solution, $\phi = \sin\omega t, \varphi = \cos\omega t$. The boundary value at a specified time $t_0$ is defined as $\phi_0 = \sin\omega t_0, \varphi_0 = \cos\omega t_0$. Therefore, $u_0 = A_0 \phi$ and the whole system becomes 34D.

We used the program AUTO [51] to analyze the bifurcations of periodic solutions. AUTO is based on a Newton iterative scheme to find the solutions. The continuation method is then used to find the branches of solutions. The bifurcations and stabilities of periodic solutions are identified by Floquet multipliers $\mu_i$. Let us recall here that, for a continuous differential system, there are three types of codimension-1 bifurcations: saddle-node (SN), period doubling (PD) and torus (TR) bifurcations [54]. The bifurcation associated with the appearance of $\mu_i=1$ ($\mu_i=-1$) is called a SN (PD) bifurcation; the bifurcation corresponding to the presence of conjugate multipliers $\mu_{1,2}=\exp(\pm i\theta_0)$, $0<\theta_0<\pi$, is called a torus bifurcation. An invariant torus is a quasiperiodic solution. Both PD bifurcations and breakdowns of invariant tori may induce chaos [50]. At present, most studies on the bifurcation are focusing on the low-dimensional systems. In our 34D model, there are multiple pairs of conjugate multipliers.

Moreover, the QR decomposition algorithm proposed by wolf *et al.* [52] is adopted to compute the spectra of LEs of the 33D system. There are 33 LEs: $\lambda_1, \lambda_2, \ldots \lambda_{33}$ (in descending order). $\lambda_1$ is named as the largest LE (LLE). All non-autonomous systems have at least one zero LE that corresponds to the *t*-component. Therefore, if $\lambda_1=0$, we discard it and make $\lambda_n^*=\lambda_{n+1}$ be the new spectrum. If $\lambda_1>0$, the motion is Lyapunov chaotic, and if there are more than two positive $\lambda_i$, the motion is described as Lyapunov hyperchaotic.

Let the quantity $s_\lambda = \sum \lambda_i^+$ be the sum of all the positive LEs. The strength of chaos is determined by the average ratio of the exponential divergence of neighbor orbits in chaotic attractor. Therefore, $s_\lambda$ may be used to describe the strength of chaos and the larger $s_\lambda$, the stronger the chaos is. Actually, noting that a larger $\lambda_1$ corresponds to a larger $s_\lambda$, it can be more conveniently used to characterize the chaos. Accordingly, if $\lambda_1^+ \to 0$, it is weak chaos which would behave as a quasiperiodic motion.

Finally, the Lyapunov dimension $d_{LD}$ is defined as $d_{LD} = j - \sum_{i=1}^{j} \lambda_i / \lambda_{j+1}$, where $j$ is the maximum value of $i$ that makes $\sum \lambda_i > 0$, *i.e.* $\sum_{i=1}^{j} \lambda_i > 0$, $\sum_{i=1}^{j+1} \lambda_i < 0$. For a high-dimensional system (>2D), $d_{LD}$ can accurately describe the fractal nature of the chaotic attractor. For a chaotic attractor, $0<d_{LD}<N_d$, where $N_d$ is the dimension of the system. For periodic and quasiperiodic motions, $d_{LD}=0$. A larger $d_{LD}$ manifests a more complex chaos.

## 4. Properties of wave propagation in the NAM models

### 4.1. Diatomic model

The bandgap properties of the diatomic NAM model have been laid out in Ref.[46]. It is shown that for this system, the generalized frequency range of the LR bandgap is $\Omega \in [1.088, 1.5]$, and the cutoff frequency of the pass band is $\Omega_c=2.251$. The transmission coefficient defined as $T_A=A_{max}/A_0$, where $A_{max}$ denotes the maximum responses amplitude. $T_A$ can be derived via three different approaches as shown in figure 3. The hardened nonlinear stiffness leads the lower bound frequency of the LR bandgap to shift upward. On the other hand, when the driving amplitude is set to $A_0=0.005$, the bandgap calculated by homotopy analysis method is $\Omega \in [1.2, 1.758]$ and $\Omega_c$ is moved to 2.277, as shown in figure 3a [46].

Actually, the pass bands of the finite LAM are composed of discrete resonances that become denser when increasing the periodic numbers. As shown in figure 3ab, although there is no damping, the appearance of nonlinearities significantly reduces and even suppresses some resonances. However, there are some differences between the numerical integral results and HAA results obtained with Newton method, especially in the second passband (*i.e.* the optical band). There are several reasons for that: (1) there is an intrinsic randomness in the chaotic responses, so $A_{max}$ manifests the extreme case but this does not mean that the system gets a steady amplitude $A_{max}$; (2) the solution of HAA is an approximately averaged result at a specified frequency and a more accurate result should require high-order harmonic components. Despite all this, HAA well depicts the nonlinear responses and the role played by the nonlinearities as well.

Different from the linear resonance with infinite amplitudes, the nonlinear modes have finite phase volumes so that the amplitudes of the 16 resonances keep bounded [46]. Increasing the amplitude reduces the wave transmissibility, and meanwhile the nonlinear upshot of suppressing resonances extends to the first passband (*i.e.* the acoustic band). This result is further demonstrated by numerical integrations.

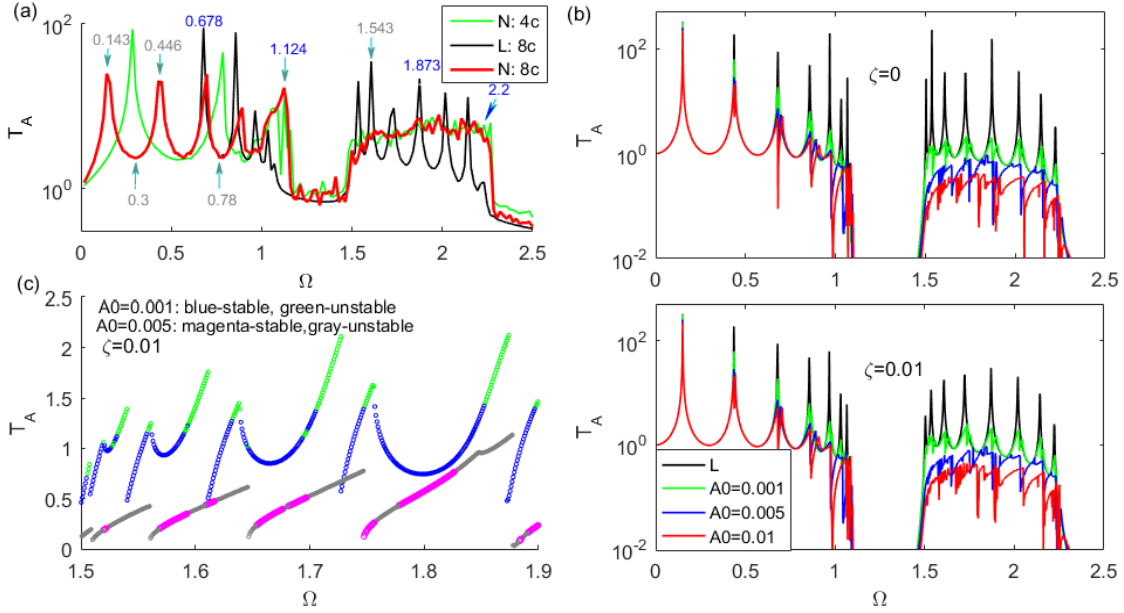

**Figure 3.** The frequency responses of diatomic LAM and NAMs. (a) maximum responses solved directly with numerical integral method; here $A_0$=0.005, $\zeta$=0, 'L' and 'N' symbolize 'linear' and 'nonlinear' models, respectively; the chain has 4 or 8 cells, where '$n$c' means there are $n$ cells in the chain; (b) the solutions found with Newton iterative method based on Eq. (14); both the model without damping ($\zeta$=0) and with a weak damping $\zeta$=0.01 are considered. (c) the response of the damped NAM in the optical branch; the solutions are solved with pseudo-arclength continuation algorithm, and the stabilities are determined with Jacobian matrix in Eq.(15); the stabilities are distinguished with different colors.

The effects on the responses of a weak damping are depicted in figure 3b. A weak damping attenuates the linear resonances but it has only little influence on the amplitudes in NAM because of the disappearance of the infinite resonances. When the nonlinearity is strong ($A_0$=0.01), $T_A$ jumps to a very low value near the resonances. The phenomenon is well explained by the pseudo-arclength continuation algorithm, as illustrated in figure 3c. When the nonlinearity is weak ($A_0$=0.001), the peaks of frequency responses of the hardened NAM bends toward the high frequency region, as what happens in a single DoF nonlinear system [55]. However, near resonances, the stable steady solutions become unstable when the frequency exceeds a threshold. Their amplitudes increase then along a branch until a stable solution appears at a certain frequency, where the amplitudes of the stable solutions switch to a lower branch. Subsequently, the amplitude increases along this branch and a "neighbor peak" with lower amplitude germinates.

When the driving amplitude increases to $A_0$=0.005, the stable and unstable solutions appear alternately. As a consequence the high amplitudes are unstable and the total frequency width of the unstable regions is broader than the width of the stable regions. However, the "neighbor peak" disappear so a curve of nonlinear resonance becomes a single trajectory bending to the right. These results illustrate the fact that, near a resonance, a high dimensional NAM has a behavior different from that of a single DoF or even of a 2DoF system [55]. Although it is difficult to determine whether the amplitude may generate a 'jump' at the catastrophe frequency, the broadband unstable solutions provide an opportunity for the chaos to appear.

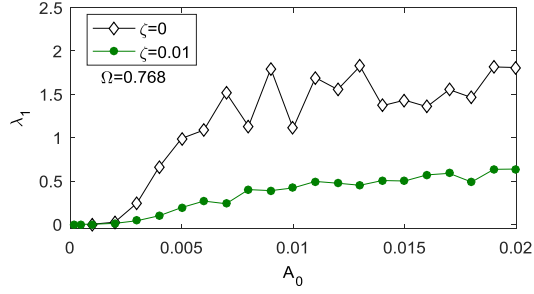

**Figure 4.** Influence of damping on the largest Lyapunov exponents ($\Omega$=0.678).

In contrast to the steady amplitude, the damping has a significant impact on the characteristics of the chaotic attractors. As shown in figure 4, in the chaotic region the LLE remarkably decreases upon damping as weak as $\zeta$=0.01. Therefore, a strong damping effect could allow a chaotic system to turn into a periodic system. Hereafter, we consider a weak damping $\zeta$=0.01 in the diatomic system.

We have chosen three illustrative frequencies $\Omega$=0.678, 1.124, $\Omega$=2.2, whose locations are labeled in figure 3a. Bifurcation diagrams and chaotic characteristics are illustrated in figure 5 for each of these frequencies.

For the conservative 34D system, there are dense TR points along the branches of periodic solutions because of the multiple pairs of conjugate multipliers, which indicates that there are extensive quasiperiodic solutions in the system. The damping does not alter the branches and stabilities of periodic solutions in the bifurcation diagram, because the linear damping $\zeta \dot{x}$ does not change the properties of the equilibriums in the time-domain algorithm (the Jacobian matrix from state space). However, the phase volumes of the damped systems shrinks in the long-term responses, which results in that the Floquet multipliers on the unit circle enter the circle. Therefore, the damping $\zeta$=0.01 reduces the number of TR points along the branches, and the PD points found in the conservative system disappear. In some frequency range, the TR point disappears too, as illustrated by figure 5c. In the main, the damping reduces the number of quasi-periodic solutions in the system.

The unstable periodic solution is an essential condition for the chaos to appear. The bifurcation diagram, mean amplitude that directly comes out from numerical integration, LEs and $d_{LD}$ in figure 5 are well consistent with each other.

In the case $\Omega$=0.678 the first unstable solution appears at $A_0$=1.867$\times$10$^{-3}$ marked by the vertical black line in figure 5. Within the subsequent domain, the stable and unstable periodic solutions appear alternately. When $A_0$>6.686$\times$10$^{-3}$, the periodic solution enters into a constantly unstable region, in which multiple branches are generated with many TR points. The terminal points of the new branches are SN points. In the interval $1\times 10^{-3}$<$A_0$<1.867$\times 10^{-3}$, $\lambda_1^+ \to 0$ and therefore the response is close to the quasiperiodic responses. Both behaviors of LEs and $d_{LD}$ attest that the periodic motion turns into chaos when the driving amplitude still increases. Actually, both $\lambda_1^+$ and $d_{LD}$ tend to increase as $A_0$ increases, meaning that the weak chaos becomes a strong chaos. However, LEs and $d_{LD}$ have local fluctuations. For the amplitude at which chaos appears, only one LE is positive, but there are quickly more than two positive LEs, *i.e.* hyperchaos occurs. Therefore, for the 33D NAM model, this is a general rule that the system is hyperchaotic when chaos occurs. Moreover, the variation laws of $d_{LD}$ are identical with $\lambda_1$ in the chaotic domain, and when strong chaos occurs $d_{LD} \to N_d$=33.

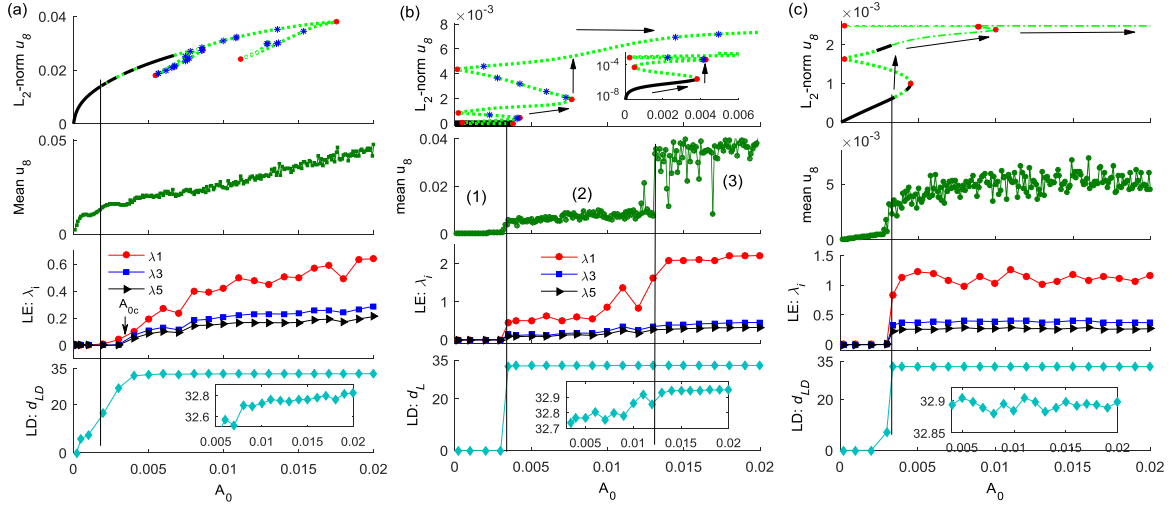

**Figure 5.** Bifurcation diagrams and chaos of diatomic model. (a) $\Omega=0.678$, (b) $\Omega=1.124$, (c) $\Omega=2.2$. $\zeta=0.01$. In each figure, the four diagrams are corresponding to the bifurcation diagram, the mean amplitude of the last linear oscillator, the Lyapunov exponents ($\lambda_1$, $\lambda_3$, $\lambda_5$) and the Lyapunov dimension $d_{LD}$, respectively. In bifurcation diagrams, the solid black (dashed green) lines represent the stable (unstable) periodic solutions, and the red dots (blue asterisks) denote the SN (TR) bifurcation point. The iconographies refer to the detail with enlarged scale. The value $L_2$-norm is defined as $\sqrt{\int_0^1 U_8(x)^2 dx}$, where $U_8(x)$ is the generalized displacement of $u_8$, and $u_8$ is the displacement of the 8th linear oscillator away from the excitation.

Combing the bifurcation diagram and LE spectrum, it is known that chaos at TR points would be induced by the breakdowns of invariant tori. A rigorous demonstration of this mechanism is based on the KAM theorem [50]. Such a demonstration falls out of the scope of this article, which rather focuses on the influence of chaos on the responses. At other unstable points, the chaos arises from period-doubling bifurcations. The waterfall plot of power spectra illustrates this process, as shown in figure 6. When $A_0 \to 0$, we are dealing with a 1:1 resonance. In this situation, the energy gets localized at the driving frequency $\Omega_e$. Further increasing $A_0$ leads to odd harmonic waves $3\Omega_e$ and then $5\Omega_e$ to appear, so the propagating waves become quasiperiodic. At the critical amplitude $A_{0c}$, the propagating wave cascades into chaos. $A_{0c}$ has the same behavior as LEs. In the route toward chaos, energy localization switches to energy dispersion and therefore the energy is pumped [56] and spread within a broad high frequency passband and even within the stop bands.

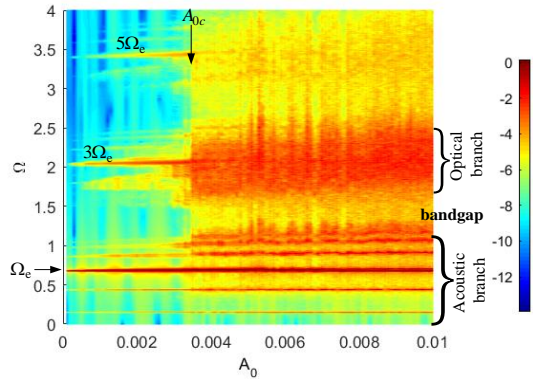

**Figure 6.** 3D waterfall plot of power spectrum density (PSD). The excitation frequency is $\Omega_e=0.678$. The different colors correspond to the value $\log_{10}(PSD)$.

If the NAM is excited by the wave at the bottom of the LR bandgap, the periodic solution exhibits multiple folded branches that induce multiple jumps, as shown by the case $\Omega=1.124$. There are still some discrete TR points on the unstable branches. The SN bifurcation points act as the terminals of branches. The first SN bifurcation occurs at $A_0=3.807\times10^{-3}$. When $A_0<3.807\times10^{-3}$, the periodic solutions remain stable and their amplitude tends to zero; both LEs and $d_{LD}$ reveal that the long-term motion of the system is periodic. The amplitudes issued from numerical integration also tend to zero. This result means that the LR bandgap remains a complete stop band and that the bandgap is linearly stable in this interval. After this SN point, the periodic solutions become constantly unstable and a jump occurs at this point. Both the bifurcation diagram and the mean amplitude show that the motion jumps to a high-energy chaotic branch at this SN point. Then, as illustrated both by LEs and by $d_{LD}$, the propagation through the NAM is along a chaotic way. Meanwhile, the hardened nonlinear stiffness shifts upwards the lower boundary of the LR bandgap.

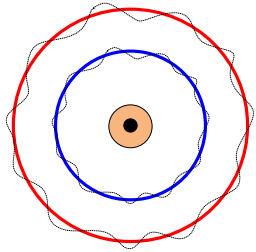

**Figure 7.** The sketch plot of quantum states in NAM. The solid circles represent the excited quantum states and the dashed curves denote the fluctuations in the neighborhood of these states. The shade circle and the black point represent the bounded quantum state.

Along this chaotic branch, the mean amplitude has almost constant value with LEs and $d_{LD}$ fluctuating in a neighborhood of the corresponding constants. Another SN bifurcation point appears at $A_0=7.676\times10^{-3}$, where the motion jumps to a higher chaotic orbit. However, this point does not fit well the long-term motion, as attested by the jumps of the mean amplitude to a higher-energy chaotic orbit at $A_0=0.013$. Like the mean amplitude, LEs and $d_{LD}$ keep almost constant. These results show that the system has multiple approximately constant states in this frequency range, and that only discrete values of these states can be achieved: the system behaves as a "quantum object", as sketched in figure 7. Therefore, when manipulating waves in the bandgaps via amplitudes, the system can switch suddenly between the stop state (*i.e.* the bounded state) and the propagation states (*i.e.* the excited quantum states). These characteristics may be helpful to realize acoustic devices with small dimensions, as for example acoustic diodes or switches. In figure 5b, it should be noted that, at some value of $A_0$, the amplitude may jump from the highest obit (the third one) to a lower one. However, these jumps are unstable and a small perturbation may stimulate the jump back to the highest orbit.

The case $\Omega=2.2$ is similar to the case $\Omega=1.124$ except that the motion turns from periodic into chaotic for only one jump at the first unstable periodic solution. The trend in bifurcation diagram agrees well with the mean amplitude variations. After the jump, the amplitude at first slowly increases and then keeps constant as do both LEs and $d_{LD}$. This suggests a mechanism to suppress elastic waves in strongly NAM: strengthening the nonlinearity decreases the wave transmissibility.

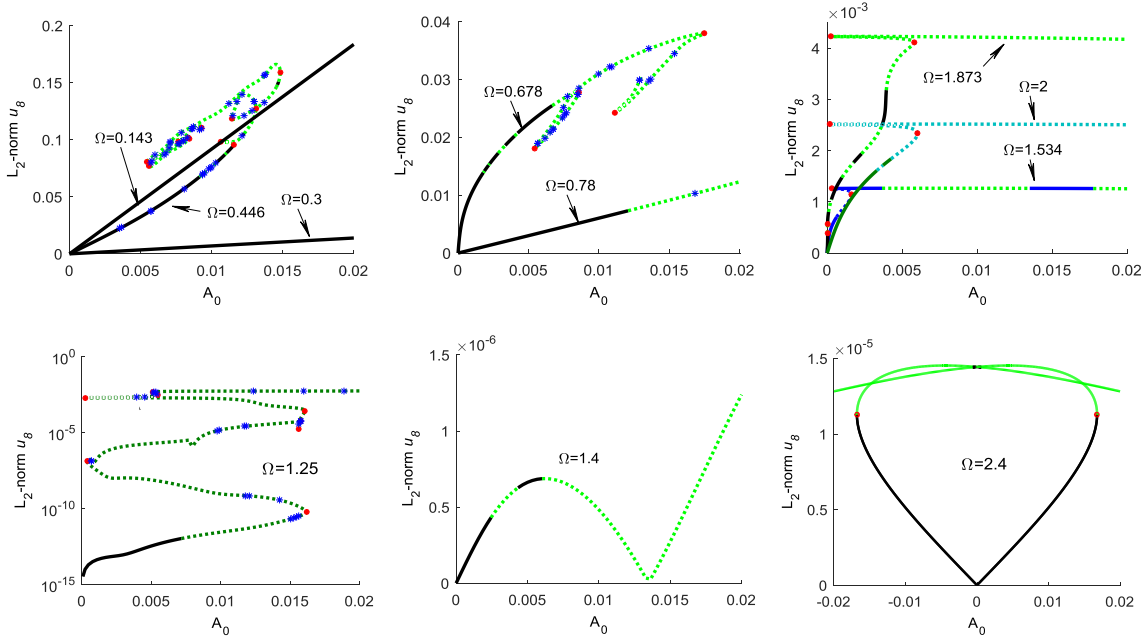

**Figure 8.** Bifurcation diagrams of diatomic model under different frequencies. The frequencies are labeled in the corresponding plots. The locations of these frequencies are marked in figure 3a. Other specifications are identical with figure 5.

More bifurcation diagrams at other frequencies are depicted in figure 8. According to the statements above, it can be anticipated that the chaos would occur at the first unstable periodic solutions. In the examined amplitude range, neither bifurcations nor unstable solutions occur on the low-frequency periodic orbits ($\Omega \leq 0.3$) and the amplitude linearly increases with driving amplitude, *i.e.* the transmissibility is constant and the linear regime is preserved. At higher frequency the motion turns to weak chaos that behaves as a quasiperiodic wave, which explains why the responses of NAM in the low-frequency parts of the acoustic branch are similar to the ones of LAM. The transmission coefficient shows this feature (see above and figure 3). In the acoustic branch, the strong chaos can be observed at the frequencies near the LR bandgap.

Bifurcation diagrams in the optical branch have similar characteristics: firstly, the amplitudes increase along an orbit where both periodic and chaotic behaviors may occur. However, when the motions jump to a higher orbit at SN point, the amplitudes of the periodic solutions do not increase with the input wave. Instead, they remain constant or even they decrease.

In the bandgap, the periodic orbits at $\Omega=1.25$ feature folding and multiple jumps, as it is the case for $\Omega=1.124$. However, a larger critical amplitude is required to force the wave into chaos. The reason is that the influences of the nonlinearity on the bandgap spread from low to high frequencies. This is further demonstrated by the bifurcation diagram at $\Omega=1.4$: although there are unstable solutions, the amplitudes are so small that only subharmonic and superharmonic waves can propagate and this frequency interval keeps its ability to reflect the incident waves. In the high-frequency stop band, as shown by the bifurcation diagram at $\Omega=2.4$, much stronger nonlinearity is needed to generate unstable solutions.

Based on all these results, the band structure of a strongly diatomic NAM is sketched in figure 9.

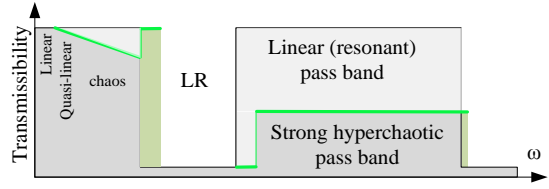

**Figure 9.** The sketch of the band structure of diatomic NAM model.

### 4.2. Tetratomic model

The dispersion relations of this model were elaborated in Ref. [47]. In the linear regime, there are three bandgaps for this metamaterial: a LR bandgap and two Bragg (BG) bandgaps. The frequency ranges of the linear bandgaps are: LR, [0.3065, 0.4195]; BG1, [1, 1.078]; BG2, [1.732, 1.777]. It is shown in Ref. [47] that, in the nonlinear regime, BG1 and BG2 are not altered if the amplitude $A_0$=0.01, but the LR bandgap is shifted to [0.42, 0.5807]. The cutoff frequency of the passband is $\Omega$=2.

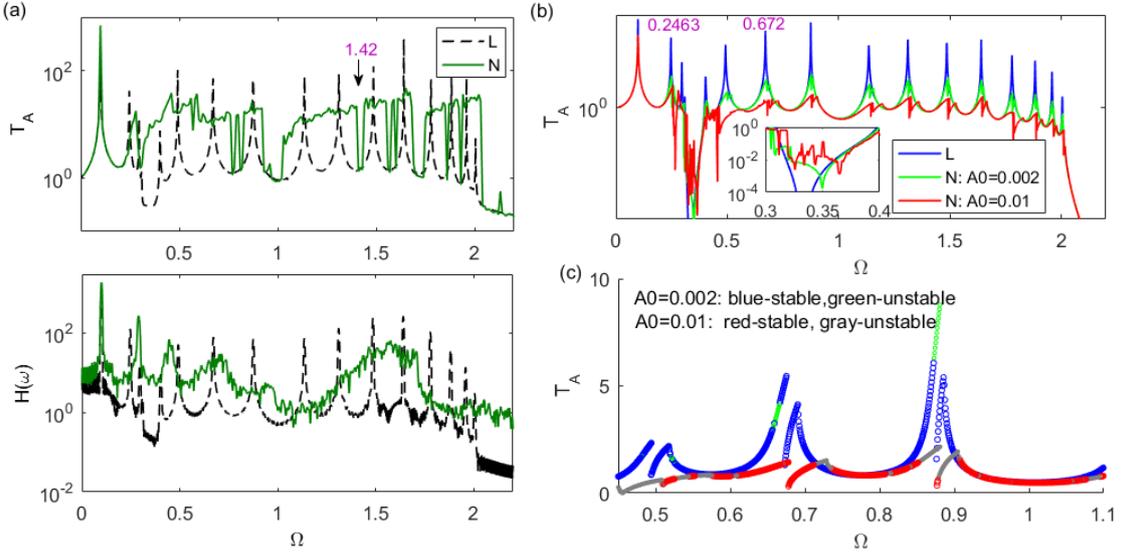

**Figure 10.** The frequency responses of tetratomic LAM and NAMs. $\zeta$=0.02 (a) $T_A$ and H($\omega$) solved directly with numerical integral method; here $A_0$=0.01, 'L' and 'N' symbolize 'linear' and 'nonlinear' models, respectively; (b) the solutions are found with the Newton iterative method; (c) the solutions are solved with pseudo-arclength continuation algorithm; the stabilities are distinguished with different colors.

We have considered a weak damping $\zeta$=0.02 in the tetratomic NAM that comprises four cells. $T_A$ computed with different algorithms, and the transfer function H($\omega$) upon sine-sweep excitations [47] are shown in figure 10. For LAM model, thanks to of its strong sensibility to localization properties, only LR bandgap is clearly evidenced by the 4-cells chain. In contrast, more cells are needed for the BG bandgaps to open up. With regard to the nonlinear model, it allows for a fair position of the BG bandgaps. As it is the case with the diatomic model, the tetratomic NAM also proves that the broadband elastic waves are suppressed. The results from HAA establish that a strong nonlinearity causes the wave transmissibility to decrease. However, the elastic waves can still propagate in the nonlinear LR bandgap. The continuation method emphasizes the surprising nonlinear modes that compose the unstable branch and a neighbor peak. Overall, the behaviors are similar to that of diatomic NAM. However, in the amplitude range $A_0 \leq 0.01$, the nonlinear modes with only one branch are not found. In the discrete models, in conjunction with the results of the diatomic NAM, we can conclude that the hardened NAM actually influences all the passbands that are higher than the 'nonlinear LR bandgap', including a small region below and near this bandgap.

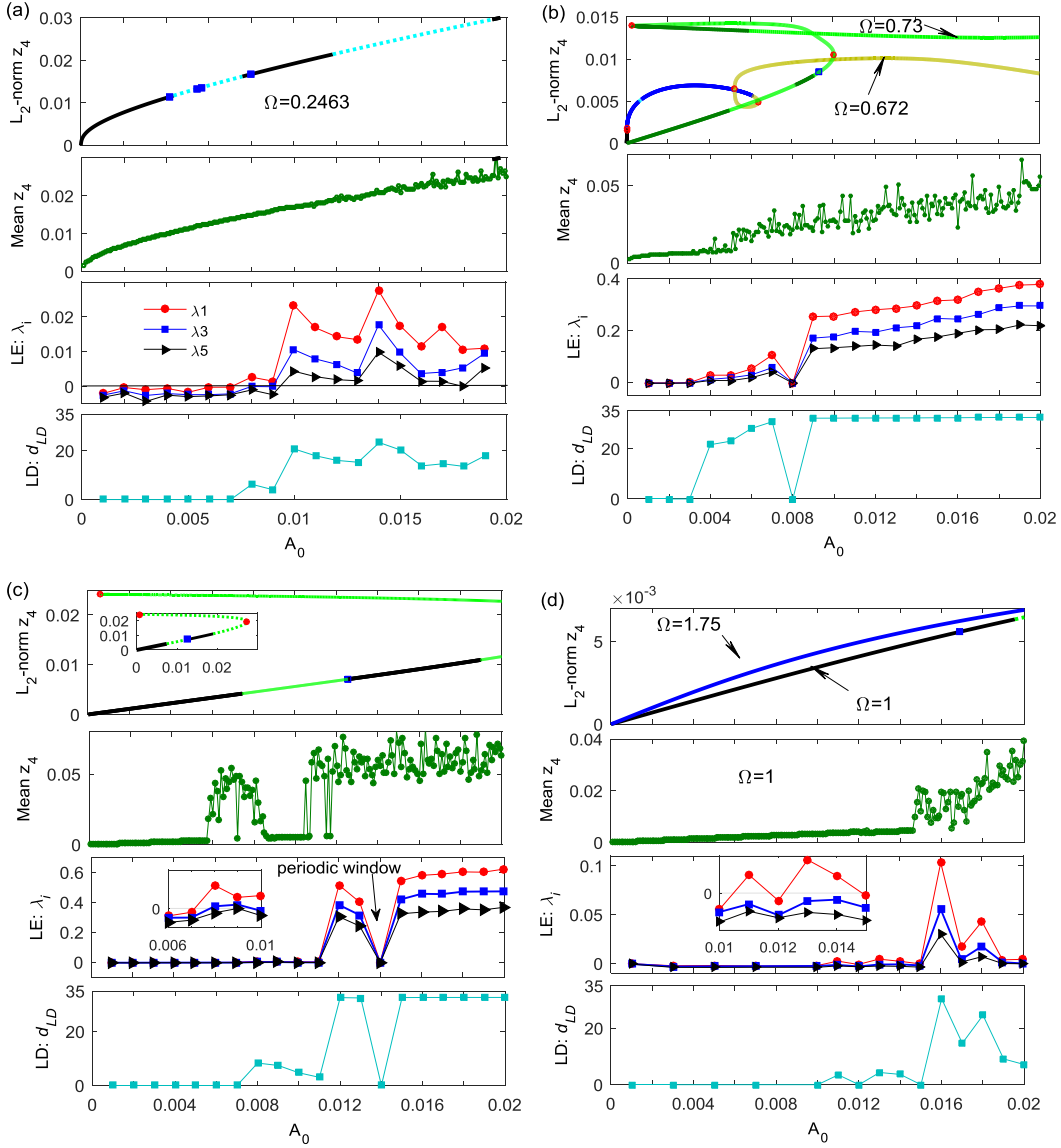

**Figure 11.** Bifurcation diagrams and chaos of tetratomic model. (a) $\Omega=0.2463$, (b) $\Omega=0.672$, (c) $\Omega=1.42$, (d) $\Omega=1$. Other specifications of these figures are same with figure 5. $z_4$ denotes the displacement of the last linear oscillator.

We have further studied bifurcation and chaos to better understand the wave propagation properties in the system. To this end, we firstly have investigated the wave propagation in the pass bands. We have considered three frequencies $\Omega=0.2463, 0.672, 1.42$ located in the 1st, 2nd, 3rd pass bands respectively, (see figure 10). The properties of the fourth passband are similar with the third one. The bifurcation diagrams, mean amplitudes, LEs and LDs are shown in figure 11a-c. Comparing these results, it can be seen that increasing the driving amplitude stimulates the motion into chaos. Moreover, there are differences between the chaotic critical points in bifurcation diagrams and LEs which are caused by: (i) the damping that greatly influences the long-term motion; (ii) the different principles implemented by the two algorithms that lead to different critical points.

In the case $\Omega=0.2463$, the bifurcation diagram is consistent with the mean amplitude: between $0.006<A_0<0.009$ the general chaos (only one positive LE) arises, but the motion turns into hyperchaos when $A_0$ further increases. However, $\lambda_1^+ <0.03$ and $d_{LD} <<33$, so that the wave undergoes a weak hyperchaos whose behavior is similar to the quasiperiodic one. Therefore, it can be deduced that the low-frequency waves in the first passband are quasiperiodic or weakly chaotic.

In fact, the propagation of the waves in the subsequent three passbands are similar except for some local differences. Along with the increasing driving amplitude, the variation law of the wave state in whole is "periodic → weakly chaotic → strongly hyper-chaotic". However, in these three passbands, the LLEs $\lambda_1$ for the strong chaos are 10 times larger than that calculated for the weak chaos in the first passband and $d_{LD}$→33. The mechanisms that suppress resonances in the passbands are identical to the ones observed with the diatomic NAM: chaos induced by period-doubling bifurcation; the wave amplitudes along the chaotic branches slowly increase, remain constant or even decrease with $A_0$. Moreover, as illustrated in figure 11(b,c), periodic windows in the chaotic region are observed, which means that LEs and LDs do not monotonously vary.

Close to $\Omega$=1.42, $T_A$ is equal to the transmission coefficient for the corresponding LAM. In fact, there are other specific frequency domains. The comparison with the results obtained with the 10-cells model [47] indicates that these domains relate to the density of the linear modes that is determined by the length of chain. In the 4-cells model, this domain is a non-resonant region. The periodic solutions for $\Omega$=1.42 have two branches: a low-energy one and a high-energy one. Along the low-energy branch, there is an unstable domain in between two stable domains. If the motion had been periodic for $A_0$≤0.01, the mean displacement of the last oscillator should have been linear against $A_0$. Instead, both the mean amplitude and LEs indicate that the system enters a weak chaos regime in the range 0.006<$A_0$<0.009. However, this weakly chaotic motion is different from that described above. Here, the oscillators jump to a high-energy unstable orbit. In the subsequent range 0.009<$A_0$≤0.011, the system jumps back to the low-energy weakly chaotic state, featuring a quasiperiodic behavior. Therefore, when the amplitude is $A_0$=0.01, the linear and nonlinear transmissibilities near $\Omega$=1.42 are equal. Further increasing the driving amplitude makes the system to jump to the strongly hyper-chaotic state, featuring constant amplitude and high-energy, except in a narrow window around $A_0$=0.014 where the motion gets periodic. . However, in this small interval, amplitudes still are in the high-energy orbit.

The jumps between low-energy and high-energy orbits explains why the waves are amplified under non-resonant conditions. This amplification scheme can be applied to design a broadband wave amplifier. For finite LAM, the waves can be amplified at the discrete resonant frequencies in the passbands. In contrast, a four-cell NAM can allow for a homogeneously amplification in a broad passband.

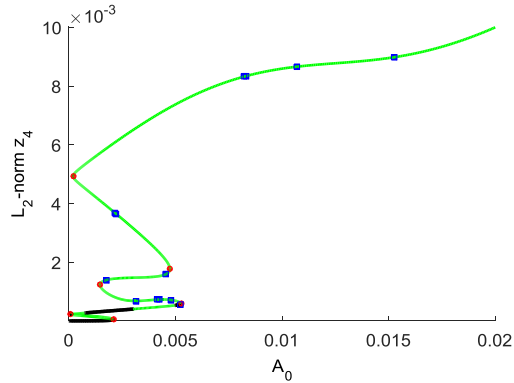

**Figure 12.** Bifurcation diagrams at frequency $\Omega$=0.35.

Another question we have addressed is: why is the nonlinear LR bandgap less efficient than other bandgaps to suppress the elastic waves in NAM? To answer this question, we have set the frequencies to $\Omega$=0.35 and $\Omega$=1. As shown in figure 12, as for the diatomic model, there are multiple bifurcations and branches. However, the stable periodic solutions disappear for an amplitude as low as $A_0$=0.003 because of the strong nonlinearity and therefore this bandgap closes at a certain driving amplitude and is replaced by a high energy chaotic orbit, as it is the case in the diatomic model. In contrast, for the BG bandgaps away from this nonlinear bandgap, much higher amplitudes are needed to get the unstable periodic solution. In the example illustrated in figure 11d, when $\Omega=1$ in the first BG bandgap, a periodic motion is observed and the chaos is weak ($\lambda_1^+ <0.04$) if the amplitude $A_0<0.01$, as for the quasi-linear state in the interval $A_0=0.01\sim0.015$. It is only as $A_0>0.015$, that the motion is attracted to the high energy hyper-chaotic state. Combing the mechanisms in diatomic model, one concludes that the nonlinear LR bandgap has weaker ability to reflect the incident waves when it jumps to the high-energy orbits.

## 5. Conclusions

The propagation of waves in the nonlinear acoustic metamaterials (NAMs) is fundamentally different from that in the conventional LAMs. However, those features are still not fully understood. In this work we investigate the elastic wave propagations in the 1D diatomic and tetratomic NAM models. We further demonstrate that nonlinear effects can greatly suppress elastic waves in broad frequency ranges. The nonlinear wave behaviors, band structures, the bifurcations and the chaos are studied to demonstrate the novel mechanisms that can manipulate wave propagations.

HAA combined with the pseudo-arclength continuation are employed to calculate the frequency responses and nonlinear modes. Bifurcations and chaos are analyzed using both the continuation algorithm and the spectra of LEs (and $d_{LD}$). Our results show that the nonlinear resonances feature multiple branches with unstable peaks and a "neighbor peak". We further have demonstrated that the period doubling process leads the motion from being periodic to chaos and damping has a significant influence on the characteristics of chaotic attractors. Moreover, due to dispersion process in chaotic regime, the localized energy spreads in the broadband high frequency passbands and even in the stop bands. The trend for the positive LEs and $d_{LD}$ is to increase with the amplitude so the weak chaos (or hyperchaos) turns into strong hyperchaos.

Band structures of both diatomic and tetratomic NAMs are studied. The bifurcations and unstable solutions do not occur on the low-frequency periodic orbits, so the responses in these regions are similar to the ones of LAMs. Actually, the hardened NAM influences all the passbands that are higher than the nonlinear LR bandgap, including a small region below this bandgap.

In the nonlinear LR bandgap, we put into evidence "quantum" behaviors because of jumping bifurcations between low-energy and high-energy orbits, whose propagation states (excited states) and stop state (bounded state) have discrete characteristics and switch suddenly. This behavior also explained why the nonlinear LR gets weaker ability to reflect the incident waves. Moreover, jumps in passbands would amplify non-resonant waves.

This work provides an important theoretical base for the understandings and applications of NAMs.

## Acknowledgements

This research was funded by the National Nature Science Foundation of China (Project Nos. 51405502 and 51275519).

## References

[1]. Kushwaha M S, Halevi P, Dobrzynski L, Djafari-Rouhani B 1993 Phys. Rev. Lett. **71** 2022–2025.
[2]. Cummer S A, Christensen J and A Alù A 2016 Nat. Rev. Mater. **1** 16001
[3]. Wang G, Wen X, Wen J, Shao L, Liu Y 2004 Phys. Rev. Lett. **93** 154302
[4]. Liu Z, Zhang X, Mao Y, Zhu Y Y, Yang Z, Chan C T and Sheng P 2000 Science **289** 1734
[5]. Xiao Y, Wen J, and Wen X 2012 New J. Phys. **14** 402380

[6]. Molerón M, Serra-Garcia M and Daraio C 2016 New J. Phys. **18** 033003
[7]. Ma G and Sheng P 2016 Sci. Adv. **2** e1501595
[8]. Rupin M, Lemoult F, Lerosey G and Roux P 2014 Phys. Rev. Lett. **112** 234301
[9]. Moiseyenko R P, Pennec Y, Marchal R, Bonello B, and Djafari-Rouhani B 2014 Phys. Rev. B **90** 134307
[10]. Huang H H and Sun C T 2009 New J. Phys. 11013003 (15pp)
[11]. Ma G, Min Y, Xiao S, Yang Z, and Sheng P 2014 Nature Mater. **13** 873
[12]. Zhang H, Xiao Y, Wen J, Yu D and Wen X 2016 Appl. Phys. Lett. **108** 141902
[13]. Zhao J, Bonello B and Boyko O 2016 Phys. Rev. B **93** 174306
[14]. Lapine M, Shadrivov IIya V and Kivshar Y S 2014 Rev. Mod. Phys. **86**(3) 1093
[15]. Leitenstorfer A, Nelson K A, Reimann K and Tanaka K 2014 New J. Phys. **16** 045016
[16]. Grady N K, Perkins Jr B G, Hwang H Y, Brandt N C, Torchinsky D, Singh R, Yan L, Trugman D, Trugman S A, Jia Q X, Taylor A J, Nelson K A and Chen H-T 2013 New J. Phys. 15 105016
[17]. Vakakis A F 1992 Acta Mech. **95** 197
[18]. Nesterenko V F 2001 *Dynamics of Heterogeneous Materials*. (Springer, New York)
[19]. Daraio C, Nesterenko V F, Herbold E B and Jin S 2006 Phys. Rev. E **73** 026610
[20]. Fang X, Zhang C H, Chen X, Wang Y S and Tan Y Y 2015 Acta Mech. **226** 1657
[21]. Nadkarni N, Arrieta A F, Chong C, Kochmann D M, and Daraio C 2016 Phys. Rev. Lett. **116** 244501
[22]. Herbold E B, Kim J, Nesterenko V F, Wang S Y, Daraio C 2009 Acta Mech. **205** 85–103
[23]. Boechler N, Theocharis G, Job S, Kevrekidis P G, Porter M A and Daraio C 2010 Phys. Rev. Lett **104** 244302
[24]. Liang L, Yuan B and Cheng J C 2009 Phys. Rev. Lett **103** 104301
[25]. Liang B, Guo X S, Tu J, Zhang D and Cheng J C 2010 Nature Mater. **9** 989-992
[26]. Boechler N, Theocharis G and Daraio C 2011 Nature Mater. **10** 665
[27]. Donahue C M, Anzel P W J, Bonanomi L, Keller T A and Daraio C 2014 Appl. Phys. Lett. **104** 014103
[28]. Zheludev N I 2010 Science **328** 582
[29]. Brunet T, J. Leng J and Olivier M-M 2013 Science **342** 323
[30]. Khajehtourian R and Hussein M I 2014 AIP Advances **4** 124308
[31]. Ganesh R and Gonella S 2015 Phys. Rev. Lett. **114** 054302
[32]. Hussein M I, Leamy M J, Ruzzene M 2014 App. Mech. Rev. **66** 040802
[33]. Manktelow K L, Leamy M J and Ruzzene M 2011 Nonlinear Dyn. **63** 193
[34]. Scalora M, Bloemer M J, Manka A S, Dowling J P, Bowden C M, Viswanathan R and Haus J W 1997 Phys. Rev. A **56** 3166
[35]. Meurer T, Qu J, and Jacobs L 2002 Int. J. Solids Struct. **39** 5585
[36]. Kim E, Li F, Chong C, Theocharis G, Yang J, and Kevrekidis P G 2015 Phys. Rev. Lett. **114** 118002
[37]. Wang S Y and Nesterenko V F, 2015 Phys. Rev. E **91** 062211
[38]. Pichard H, Duclos A, Groby J-P, Tournat V, Zheng L and Gusev V E 2016 Phys. Rev. E **93** 023008
[39]. Lydon J, Theocharis G and Daraio C 2015 Phys. Rev. E **91** 023208
[40]. Narisetti R K, Ruzzene M and Leamy M J 2011 J. Vib. Acoust. **133** 061020
[41]. Narisetti R K, Ruzzene M and Leamy M J 2012 Wave Motion **49** 394
[42]. Manktelow K, Leamy M J and Ruzzene 2013 J. Mech. Phys. Solids **61** 2433
[43]. Yousefzadeh B and Phani A S 2015 J. Sound. Vib. **354** 180
[44]. Bernard B P, Mazzoleni M J, Garraud N, Arnold D P and Mann B P 2014 J.Appl. Phys. **116** 084904
[45]. Xu Y and Nesterenko V F 2014 Phil. Trans. R. Soc. A **372** 20130186
[46]. Fang X, Wen J, Yin J Yu D and Xiao Y 2016 Phys. Rev. E **94** 052206
[47]. Fang X, Wen J, Yin J and Yu D 2016 AIP Advances **6** 121706
[48]. Cheung Y K, Chen S H and Lau S L 1990 J. Sound. Vib. **140**(2) 273-286
[49]. Xu D L, Zhang H C, Lu C, Qi E R, Tian C and Wu Y S 2014 Phys. Rev. E **89** 042906
[50]. Kuznetsov Yuri A 1998 *Elements of Applied Bifurcation Theory*, 2nd Edition. (Springer, New York)
[51]. Doedel E J and Oldeman B E AUTO-07P 2012 Continuation and bifurcation software for ordinary differential equations, available at http://indy.cs.concordia.ca/auto
[52]. Wolf A, Swift J B, Swinney H L and Vastano J A 1985 Physica D **16** 285-317
[53]. Eckmann J P and Ruelle D 1985 Rev. Mod. Phys. **57** 617-656
[54]. Strogatz S H, *Nonlinear dynamics and chaos*. (Westview Press, 2015).
[55]. Nayfeh A H and Mook D T, *Nonlinear oscillations*. (Wiley, New York, 1979).
[56]. Fang X, Wen J, Yin J, and Yu D 2016 Nonlinear Dynamics; doi: 10.1007/s11071-016-3220-4